\documentclass[a4paper,fleqn]{article}
\usepackage[utf8]{inputenc}
\usepackage[numbers]{natbib}
\usepackage{physics}
\usepackage{graphicx}
\usepackage{xcolor}
\usepackage[normalem]{ulem}

\usepackage{authblk}
\usepackage{amssymb}
\usepackage[font=small,labelfont=bf]{caption}
\usepackage[margin=1.0in]{geometry}
\begin{document}

%
%
%
%
%
\title{Towards an effective mass model for the quasi-1D magnesium diboride superconducting nanostructures}
\author{Wojciech Julian Pasek, Marcos H.~Degani, Marcelo Z.~Maialle}
\affil{Faculdade de Ciências Aplicadas, Universidade Estadual de Campinas,  R.~Pedro Zaccaria 1300, Limeira, SP 13484-350, Brazil}
\author{Marcio C.~de Andrade}
\affil{Cryogenic Electronics and Quantum Research Branch, Naval Information Warfare Center Pacific, San Diego, CA 92152, USA}
\date{19 January 2022}
%

\maketitle

\begin{abstract}
We developed a parabolic effective mass model with the aim to describe the in-plane constricted superconductivity in few-monolayer MgB$_2$. The model was employed to study a constricted quasi-$1$D superlattice with the simplest version of the Anderson approximation. Resonant Fano-Feshbach-like effect was found, between a miniband of a continuous normal state dispersion cutting across $E_F$ and a miniband of quasi-discrete dispersion, even though the latter is significantly energetically distanced from the Fermi level. The origin of the phenomenon was identified to be the presence of characteristic strongly localised eigenstates, due to the specific material parameters. Moreover, the mechanism leading to a "precursor" pseudogap region in a system with a non-zero phase difference is described, when the strength of the coupling is still not sufficient to open a superconducting gap, but enough to introduce electron-hole mixing into the system.
\end{abstract}

\section{Introduction}
Quantised electronic states in superconducting (SC) nanostructures have great potential for applications in novel devices, such as for field sensors \cite{collienne2021nb} or for quantum computation and Majorana states platforms \cite{hasan2010colloquium}. In particular, the density of states (DOS) close to the Fermi energy, where the SC pairing interactions take place, is modified by the quantisation. This is commonly known as the "shape resonance” effect \cite{valentinis2016rise}, meaning that the system’s geometry can cause novel coherent quantum phenomena, potentially enhancing the SC pairing energy and the critical temperature T$_c$.

In the nanostrutures, it is possible to have carrier movement quantised along some specific spatial directions while still existing free motion constrained in the complementary directions. The constrained free motion can be described by a set of energy dispersions for each quantised state, which are then referred to as “subbands''. For instance, planar systems have two-dimensional (2D) subbands, as realised in SC nanofilms, planar layered materials (e.g.~MgB$_2$), and on the surface of topologically protected bulk materials. Linear systems exhibit 1D dispersion for the subbands, as e.g.~for thin nanowires. A hybrid system, quasi-1D (Q1D), can be thought of as a nanoribbon \cite{guan2010sample}. This scenario allows for the carrier movement to occur in several subband channels, each related to different quantised states. Multichannel systems show interesting properties arising from the interplay between SC coupling of states in the same or in different condensate components (similarly for subbands) – e.g.~gap anisotropy \cite{saunderson2020gap}, pseudogap phases \cite{salasnich2019screening}, enhancement of the SC gap and the critical temperature \cite{vargas2020crossband}, interband fluctuation screening \cite{salasnich2019screening,tajima2020mechanisms}, and electronic topological transitions \cite{bianconi2005fesh}.

In the early 2000s, the observation of high-temperature superconductivity in MgB$_2$, a multiband superconductor, have ignited the interest in possible superconductivity types where many carrier bands contribute to the condensate state. For bulk MgB$_2$, two-band models have been sufficient to fully describe the SC state with the two spectral gaps associated with the $\pi$ and $\sigma$ states \cite{vargas2020crossband,jin2019topological,choi2002origin}. It has been found that an appropriated SC coupling between a deep band ($\pi$) and a shallow band ($\sigma$) can give rise to a notable increase of SC gaps \cite{guidini2014band,bianconi2004thetc}. However, shallow bands are particularly susceptible to fluctuations that destroy superconductivity. On the other hand, it was shown that these fluctuations can be effectively screened if the shallow band is appropriated coupled to deep-band states \cite{salasnich2019screening,tajima2020mechanisms}.

Q1D nanostructures potentially allow for some beneficial geometric control of the  SC properties. However, Q1D systems appear to have superconductivity suppressed due to large ﬂuctuations of the order parameter. As mentioned above, there may be a situation in which these fluctuations can be effectively screened in multiband systems. This possibility, in addition to the fact that even considerably weak coupling between multiple condensates reinforces the SC gaps, led to the prediction that Q1D systems can host enhanced SC properties \cite{saraiva2020multiband}.

When a Q1D structure has its free motion further modulated periodically in space, each subband dispersion gives rise to a set of “minibands”-- as it is often realised in semiconductor superlattices. The parameters of the modulated potential can function as additional control of the SC properties by modifying the miniband dispersion, for instance, affecting the DOS at the Fermi level \cite{bianconi1998superconductivity}.

In recent work \cite{pasek2021band}, we have investigated a Q1D system consisting of a nanoribbon periodically constricted along with its length, as shown in Fig.~\ref{fig:potential}(a). The system was modelled using NdSe$_2$ material parameters and the corresponding Bogoliubov-de Gennes (BdG) equations were solved self-consistently in the Anderson approximation. We had observed that while at some Fermi momentum point in the miniband dispersion there was a SC gap, in other momentum points the gap was closed. This pseudogap SC state could occur because the corresponding miniband had mixed symmetry for the transverse quantised states, a consequence of the asymmetry of the constrictions. In the present work, we further investigate the periodically constricted nanoribbon, however, using material parametres for a few-layer MgB$_2$. This is a much richer system because MgB$_2$ is already a multiband superconductor. Therefore, the coupling between pair states in different subbands and different minibands have now many other possibilities to occur. We work in the envelope ansatz (effective mass model) in which the multiband nature of MgB$_2$ is taken into account by fitting the material parameters. Because of the great computational complexity, we build our model ground up, including two $\sigma$ subband pairs and one $\pi$ subband. This minimal model is our platform to study how the inter-subband/miniband couplings can introduce new dynamics to the formation of superconductivity.

The SC properties are usually thought to be decided by the characteristics of the normal system at, or close to, the Fermi level. This work shows an unusual case in which a miniband far from the Fermi level, but within the limit of the Debye window, had a very strong impact on the SC states. The calculated charge densities of the  normal states of this miniband are very strongly localised, lying wholly in the central constricted section of the primitive cell (cf.~Fig.~1). This peculiar character leads to an unusually strong contribution to the order parameter, which can be considered as a "remote" Fano-Feshbach resonance \cite{salasnich2005condensate}. The purpose of the present work is to demonstrate the numerical framework used in our modelling and to explore the characteristics of SC states in the MgB$_2$ nanoribbon. In particular, the effect of the localised miniband in enhancing the superconductivity is addressed, but without studying the inter-subband couplings.

\section{Model}

\subsection{Geometry}
\begin{figure}[htp]
     \centering
     \begin{minipage}[b]{0.496\textwidth}
         \centering
         \includegraphics[width=\textwidth]{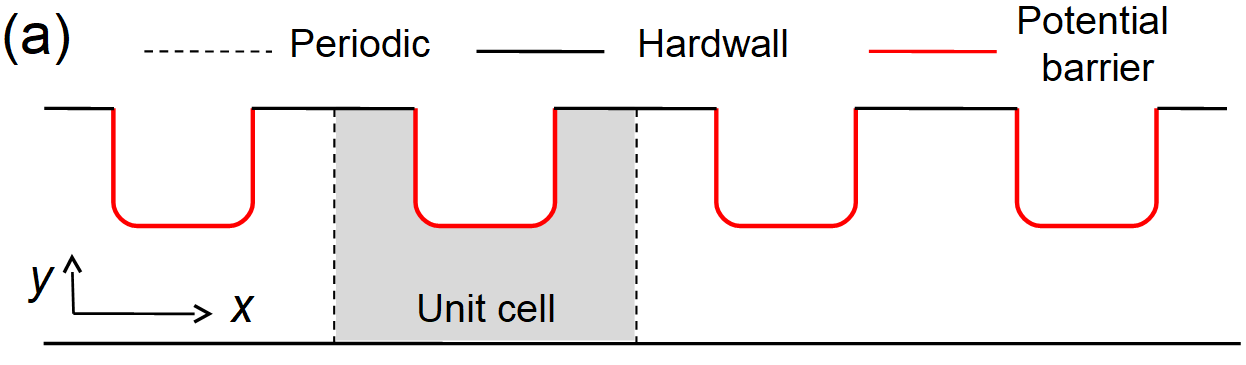}
     \end{minipage}
     \begin{minipage}[b]{0.496\textwidth}
         \centering
         \includegraphics[width=0.8\textwidth]{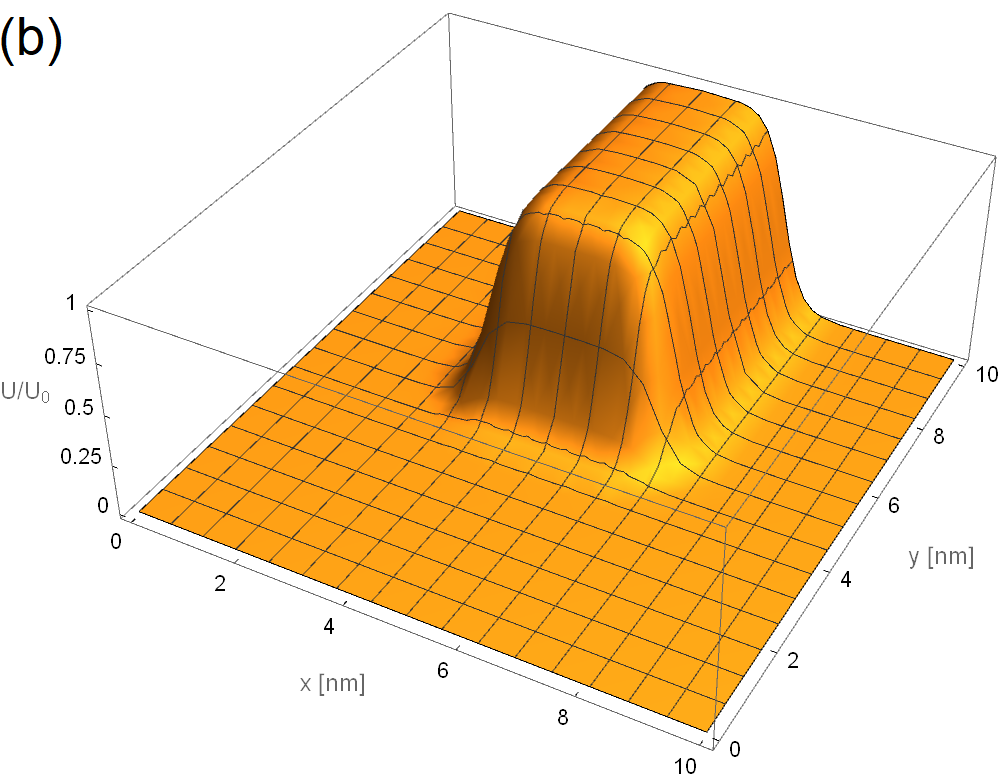}
     \end{minipage}
    \caption{(a) Schematics of the nanoribbon. The dashed lines mark a primitive cell. The types of lines indicate the corresponding boundary conditions. (b) Potential in the nanoribbon primitive cell. The relative barrier (potential energy) profile $U/U_0{\sim}0$ corresponds to the unconstricted region and $U/U_0{\sim}1$ to the barrier region.}\label{fig:potential}
\end{figure}

The material of the nanostructure is the few-layer MgB$_2$, see e.g.~\cite{bekaert2017free}. The quasi-$2$D nature of the material translates into a few subband dispersion relation near the Fermi level. A Q1D superlattice (SL) is considered, Fig.~\ref{fig:potential}, with its longitudinal $x$ axis oriented along ${\Gamma}K$ and the finite transverse $y$ direction along ${\Gamma}M$. The primitive cell of the SL has the size $L_X = L_Y = 10$ nm, with about a sixth of the cell $(L_X /3 < x < 2 L_X /3, L_Y /2 < y)$ constituting effectively a barrier. The boundary conditions are periodic in $x$ and hardwall in $y$. This geometry, for the normal state, leads to a set of SL miniband ladders, one for each of the material subbands. The $x$ and $y$ directions are not separable, which leads to significant mixing of corresponding symmetries.

\subsection{Fitting the dispersion of the material subbands}

We work in the envelope ansatz, i.e.~effective mass model. The multiband nature of MgB$_2$ is taken into account by fitting the material parameters (band edges and effective masses) for each material subband in the thickness ($z$) direction. Furthermore, because of the hexagonal crystal structure, the fit must be done separately for transverse ($y$) and longitudinal ($x$) in-plane directions, because if the $x$ axis is oriented along ${\Gamma}K$ direction, then $y$ lies along ${\Gamma}M$ direction. Starting from the data from ab initio calculations \cite{bekaert2017free}, we fit parabolic dispersions,\footnote{Specifically, we use data from \textit{Figure $5$. The calculated band structure of freestanding $6$ ML thick MgB$_2$} of the cited work.} centred at the $\Gamma$ point for the ten $\sigma$ subbands (pairs 1-5) and three $\pi$ subbands. Please note that each of the $\sigma$ pairs is degenerated at the maximum.
We are mostly interested in states whose energies lie in the relative vicinity of the Fermi level (zero energy on our energy scale). However, the eigenstates of the nanostructure do not have defined crystal $k_{x(y)}$ as in the case of the material itself and instead have a spectrum of different $k_{x(y)}$ components. Moreover, additional energy shift will come in our system from the potential energy present in the primitive cell. This means that one cannot limit the fit to only small vicinity of $E_F$ and must make a more "global" fit. In case of the $\sigma$ subbands we use: the energy shift at $\Gamma$ as well as the $k$ values at several points in the $E{\in}(-1, 0)$~eV range. It turns out that the $\sigma$ pairs can be assumed to be equidistant $[E_{\Gamma}(n) = E_{1,\Gamma} + (n-1) {\Delta}E_{\Gamma}]$. Moreover, the same band masses ($m_{lx}$, $m_{ly}$) can be used for all lower (\textit{l}) subbands in each pair, as well as ($m_{ux}$, $m_{uy}$) for the five upper (\textit{u}) subbands of each pair. This leads to a set of six parameters for the ten dispersions $[E_{1,\Gamma}, {\Delta}E_{\Gamma}, m_{lx}, m_{ly}, m_{ux}, m_{uy}]$. The $\pi$ subbands have a more complicated shape, and each needs to have the set of 3 parameters $(E_{\Gamma}, m_x , m_y)$ fitted separately. In this case the offset energies at $\Gamma$ read from \cite{bekaert2017free} cannot be used, as they spoil the shift, but we deem this irrelevant as these energies are very far away from the $E_F$ anyway. The dispersion over the $E{\in}(-0.5, 1)$~eV interval was used in the $\pi$ fit. The fitted dispersions are presented in Fig.~\ref{fig:fitted_dispersions} and the values of the fitted parameters are given in Table~\ref{tab:eff_masses}.

\begin{table}[ht]
\begin{center}
\begin{tabular}{c|c|c|c}
Quantity & Band & Direction: crystal (SL) & Value \\ \hline
$m$ & $\sigma$ lower & M ($y$) & $-0.150$ \\ \hline
$m$ & $\sigma$ upper & M ($y$) & $-0.440$ \\ \hline
$m$ & $\sigma$ lower & K ($x$) & $-0.289$ \\ \hline
$m$ & $\sigma$ upper & K ($x$) & $-0.521$ \\ \hline
$m$ & $\pi_1$ & M ($y$) & $-0.592$ \\ \hline
$m$ & $\pi_1$ & K ($x$) & $-0.726$ \\ \hline
$m$ & $\pi_2$ & M ($y$) & $-0.688$ \\ \hline
$m$ & $\pi_2$ & K ($x$) & $-0.834$ \\ \hline
$m$ & $\pi_3$ & M ($y$) & $-0.763$ \\ \hline
$m$ & $\pi_3$ & K ($x$) & $-0.963$ \\ \hline
$E_{1,\Gamma}$ & $\sigma$ & & $267.0$ meV \\ \hline
${\Delta}E_{\Gamma}$ & $\sigma$ &  & $95.2$ meV \\ \hline
$E_{\Gamma}$ & $\pi_1$ & & $5.689$ eV \\ \hline
$E_{\Gamma}$ & $\pi_2$ & & $5.587$ eV \\ \hline
$E_{\Gamma}$ & $\pi_3$ & & $5.990$ eV
\end{tabular}
\caption{The values of effective masses and energy offsets fitted to the data of Fig.~$5$ of \cite{bekaert2017free}.}\label{tab:eff_masses}
\end{center}
\end{table}

\subsection{Single particle model}

\begin{figure}[htp]
    \includegraphics[width=0.496\textwidth]{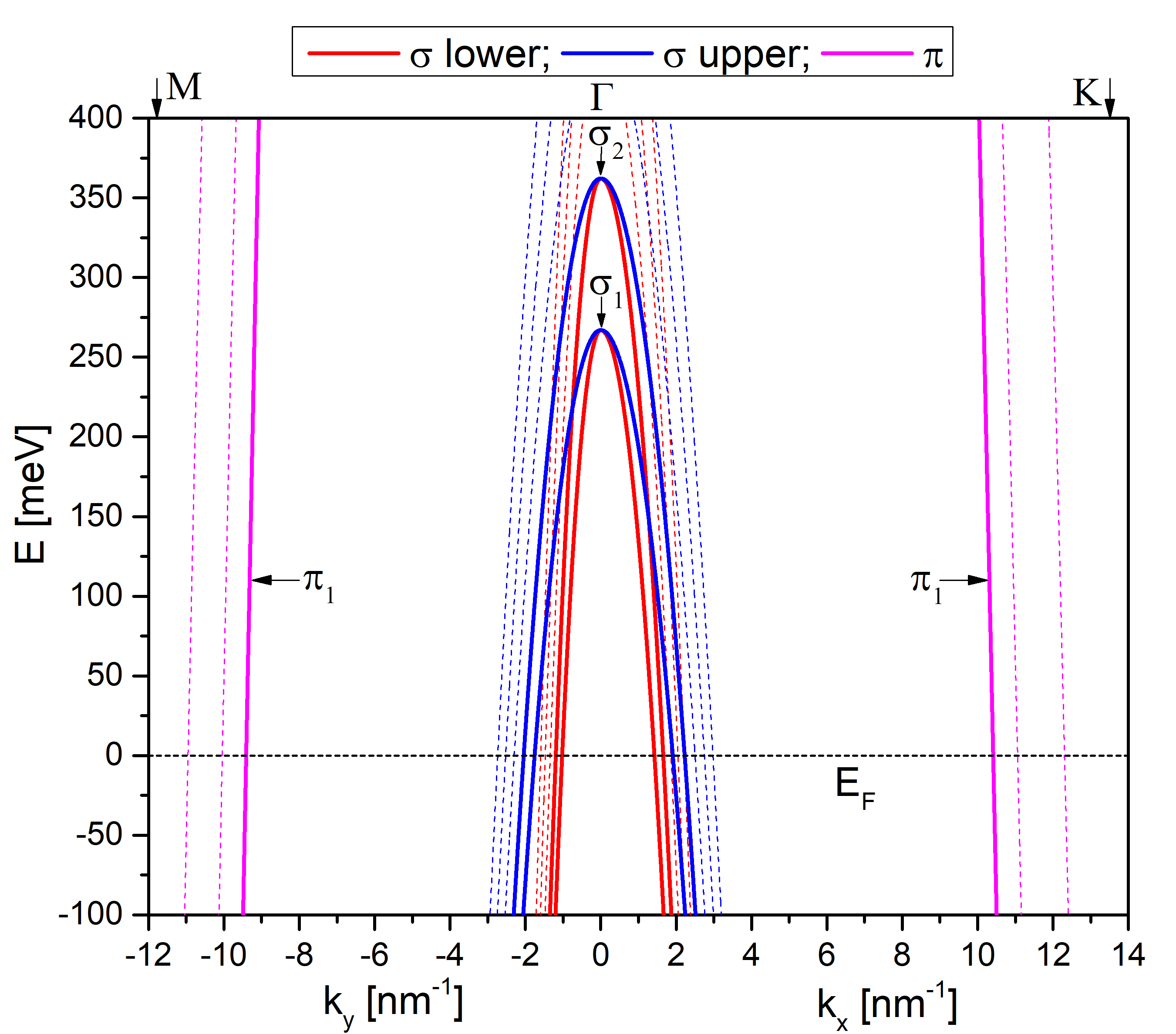}
    \caption{Parabolic dispersion as fitted in our model. Thick solid lines show the subbands taken into account in the minimum model presently under investigation.}\label{fig:fitted_dispersions}
\end{figure}

The material parameters are used in the Hamiltonian of the kinetic normal state
\begin{equation}
T = E_{\Gamma} + \frac{k_x^2}{2 m_x} + \frac{k_y^2}{2 m_y}.
\end{equation}
Here, the effective masses have negative values, which results in hole-like dispersions, see Fig.~\ref{fig:fitted_dispersions}. The periodic constriction in the nanoribbon is introduced by a smooth potential energy $U$, equal to zero in the unconstricted region and prohibitively large ($U_0 = 1$ keV) in the barrier. The exact shape of the barrier is defined as
\begin{equation}
	U = U_0~f_X\left(x,L_X,s_X\right)~f_Y\left(y,L_Y,s_Y\right),
\label{eq:potential_energy}\end{equation}
where
\small\begin{equation}
\begin{split}
	f_X\left(x,L_X,s_X\right) &= \frac{e^{\frac{L_X}{3 s_x}} \left(e^{\frac{L_X}{6 s_x}}+1\right)^2 \left(e^{\frac{L_X}{s_x}} - e^{\frac{x}{s_x}}\right) \left(e^{\frac{x}{s_x}}-1\right)}{\left(e^{\frac{L_X}{2 s_x}}-1\right)^2 \left(e^{\frac{L_X}{3 s_x}} + e^{\frac{x}{s_x}}\right) \left(e^{\frac{2 L_X}{3 s_x}} + e^{\frac{x}{s_x}}\right)}\\
	f_Y\left(y,L_Y,s_Y\right) &= \frac{e^{\frac{L_Y}{2 s_Y}} \left(e^{\frac{y}{s_Y}}-1\right)}{\left(e^{\frac{L_Y}{2 s_Y}}-1\right) \left(e^{\frac{L_Y}{2 s_Y}} + e^{\frac{y}{s_Y}}\right)}\label{eq:potential_energy_2}
\end{split}
\end{equation}
\normalsize
and is shown in Fig.~\ref{fig:potential}(b).\footnote{While $f_Y\left(y,L_Y,s_Y\right)$ and $f_X\left(x,L_X,s_X\right)$ may seem to have complicated form, in fact their origin is simple. The former is zero for $y=0$, is equal to one for $y=L_Y$ and to one half when $y=L_Y/2$. The latter is zero for $x=0$ or $x=L_X$, is equal to one for $x=L_X/2$ and to one half when $x=L_X/3$ or $x=2 L_X/3$. Starting from Fermi-Dirac kind of function, after demanding the mentioned properties, one gets the function we use.}

The SL is modelled by using the periodic boundary conditions in the longitudinal direction, leading to the Bloch-like expression $\Psi(x,y) = \Phi(x,y) \exp(i Q x)$, where $Q$ is the SL wavenumber $-\pi/L_X < Q < \pi/L_X$, and $\Phi(x,y)$ has the periodicity of the SL. In practice, we use a mesh of $n_Q = 97$ points to represent the $Q$ spectrum.

The normal state Hamiltonian $H_{ns} = T + U$ is diagonalised in a set of functions. In the longitudinal direction we use the set of states $\exp(2 i \pi n_X \frac{x}{L_X})$ and in the transverse direction $\sin(\pi n_Y \frac{y}{L_Y})$, where $-N_X \leqslant n_X \leqslant N_X$ and $1 \leqslant n_Y \leqslant N_Y$. The difference between the function sets comes from the different types of boundary conditions. The $N_X = 21$ and $N_Y = 37$ come from the envelope ansatz, specifically from the assumption that $k_x$ and $k_y$ lie inside the crystal Brillouin zone.\footnote{Note that, while the $k$-dynamic of the real crystal dispersion reverses after going from the first to the second Brillouin zone, e.g. returning to the $k=0$ value for the $\Gamma$ point of the second B.z., the parabolic fit diverges to $-\infty$.} 

\subsection{Anderson model}

The formation of the SC coupling is obtained by self-consistently solving the set of BdG equations, within the Anderson approximation. Let’s take the set of the normal state eigenenergies $\epsilon_{Q,n}$ for a single subband, where the $n$ quantum number numerates the minibands. Then, the basic Anderson model amounts to a $2{\times}2$ matrix:
\small\begin{equation}
\left(\begin{array}{cc}
    \epsilon_{Q,n} & \Delta_{Q,n,\dd{Q}}\\
    \Delta^{*}_{Q,n,\dd{Q}} & -\epsilon_{Q+\dd{Q},n}
\end{array}\right)\left( \begin{array}{c}
    U_{Q,n}\\
    V_{Q+\dd{Q},n}
\end{array}\right)
    = E_{Q,\dd{Q},n}
\left(\begin{array}{c}
    U_{Q,n}\\
    V_{Q+\dd{Q},n}
\end{array}\right)
\end{equation}\normalsize
where $U_{Q,n}$ and $V_{Q+dQ,n}$ are the electron- and hole-like coefficients, $U_{Q,n}^2 + V_{Q+dQ,n}^2 = 1$, and ${\Delta}_{Q,n,dQ}$ is the order parameter matrix element, defined as:
\small\begin{multline}
\Delta_{Q,n,\dd{Q}} = J_0 \sum_{Q^\prime} \sum_m U_{Q^\prime,m} V^{*}_{Q^\prime+\dd{Q},m} \kappa_{Q,n,Q^\prime,m,\dd{Q}} \\ \left[ 1-2 F_D\left(E_{Q^\prime,\dd{Q},m},T\right) \right] F_D\left(E_{Q^\prime,\dd{Q},m}-E_C,T_D\right)
\label{eq:Delta_element}\end{multline}\normalsize
with $F_D(E,T)$ the Fermi-Dirac function for energy $E$ and temperature $T$. In this work we set $T = 0.5$~K. The mixing of electron-like and hole-like parts of SL momenta $Q$ and $Q+dQ$, respectively, leads to the phase difference $d\phi = -dQ L_X$. This is the phase difference of the order parameter per one unit cell, that is $\Delta(x+L_X,y) = \Delta(x,y) \exp(i d\phi)$. The last term in the $\Delta_{Q,n,dQ}$ definition smoothly ($T_D = 50$~K) switches the contribution to the order parameter on, for the states lying inside the Debye window ($E<E_C=75$~meV), and off, outside. $J_0$ is the constant corresponding to the general strength of the coupling in units of energy, and the “heart” of the sum is the dimensionless four-orbital contact term
\small\begin{equation}
    \kappa_{Q,n,Q^\prime,m,\dd{Q}} = L_X L_Y \iint \Phi^{*}_{Q,n} \Phi_{Q+\dd{Q},n} \Phi_{Q^\prime,m} \Phi^{*}_{Q^\prime+\dd{Q},m} \dd{x} \dd{y}.
\end{equation}\normalsize

This usual (minimal) Anderson model allows only for direct coupling (in the sense of the electron- and hole-like mixing to form the SC state) at a given $Q$ between the same SL miniband $n$ of the same material subband. Indirectly, via the generation of the order parameter $\Delta$, it allows this inter-miniband coupling to be introduced in some state of miniband $n$ at wavevector $Q$ by the coupling happening within another miniband $m$ and/or at different wavevector $Q^\prime$. Still, any kind of coupling between the states corresponding to different material subbands is disregarded in the present work.

\section{Results}

\begin{figure*}[htp]
     \centering
     \begin{minipage}[b]{0.496\textwidth}
         \centering
         \includegraphics[width=\textwidth]{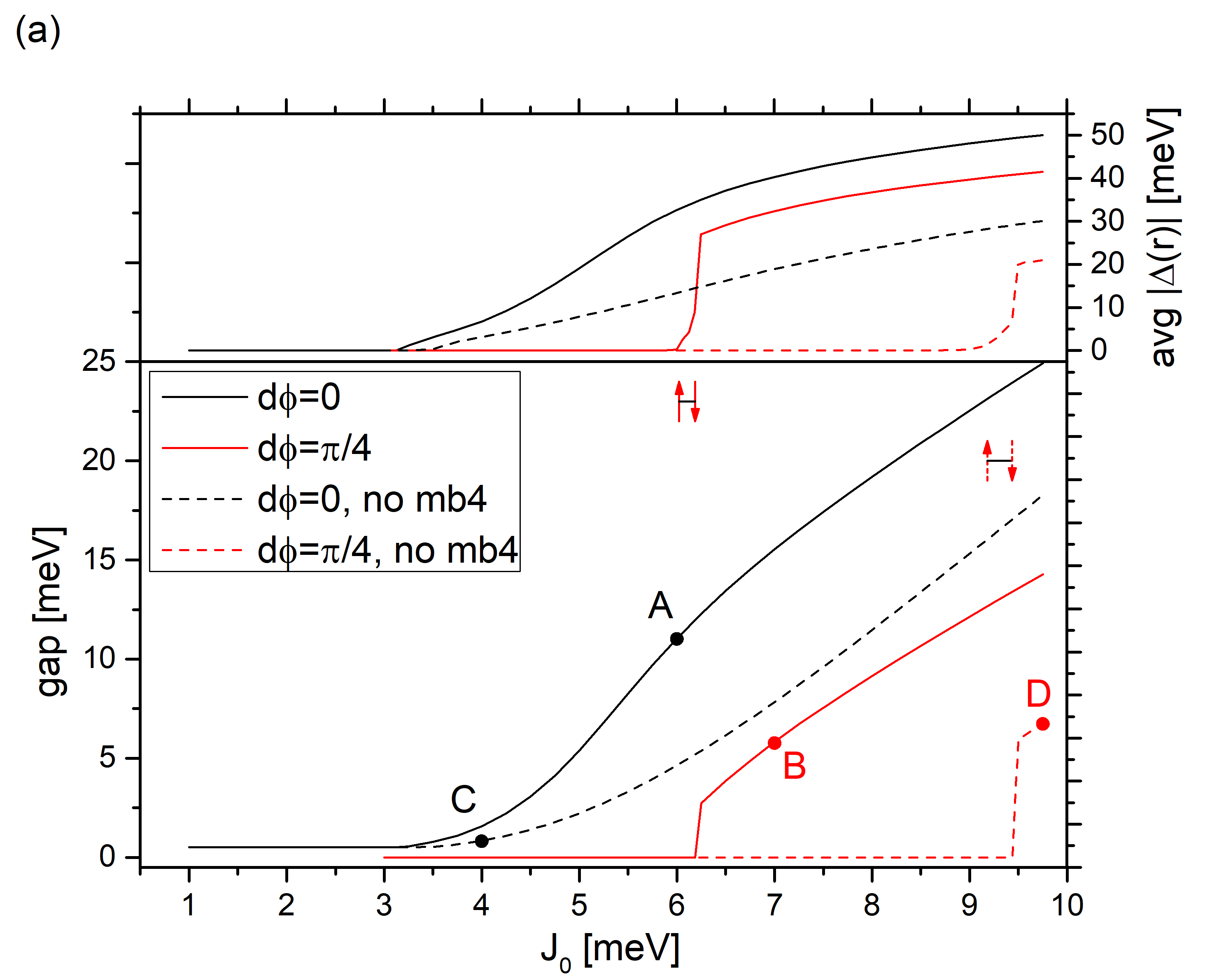}
     \end{minipage}
     \begin{minipage}[b]{0.496\textwidth}
         \centering
         \includegraphics[width=\textwidth]{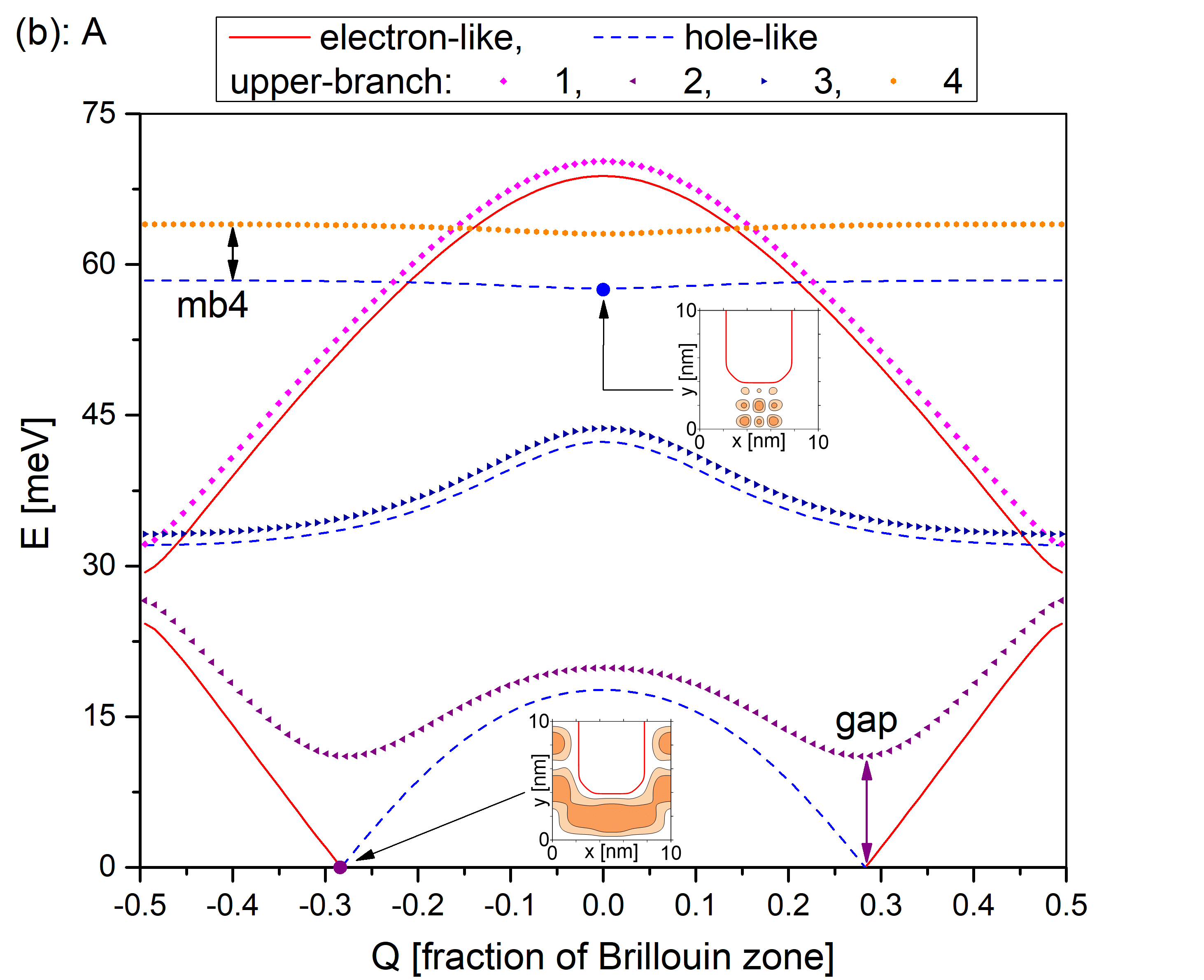}
     \end{minipage}
     \begin{minipage}[b]{0.496\textwidth}
         \centering
         \includegraphics[width=\textwidth]{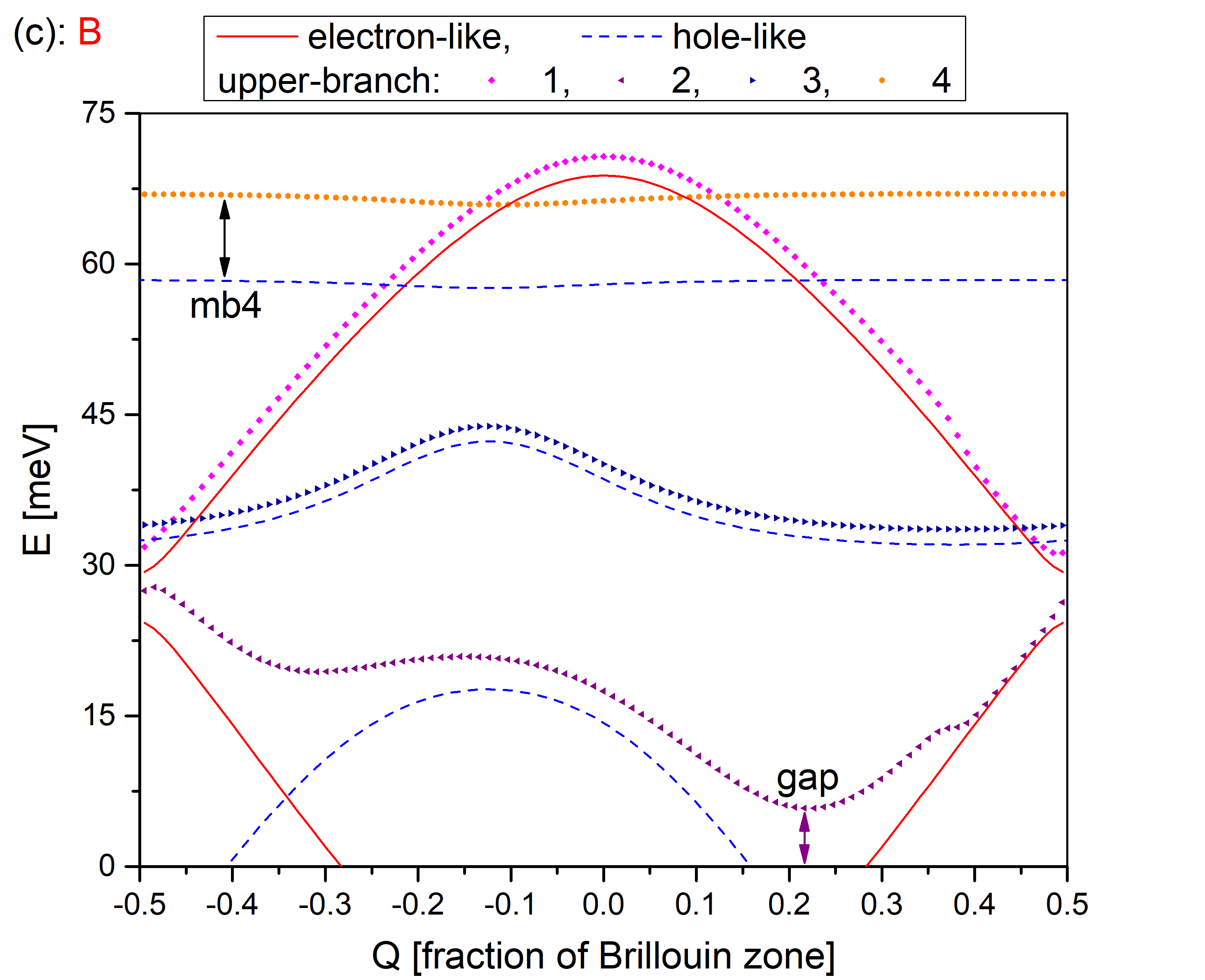}
     \end{minipage}
     \begin{minipage}[b]{0.496\textwidth}
         \centering
         \includegraphics[width=0.85\textwidth]{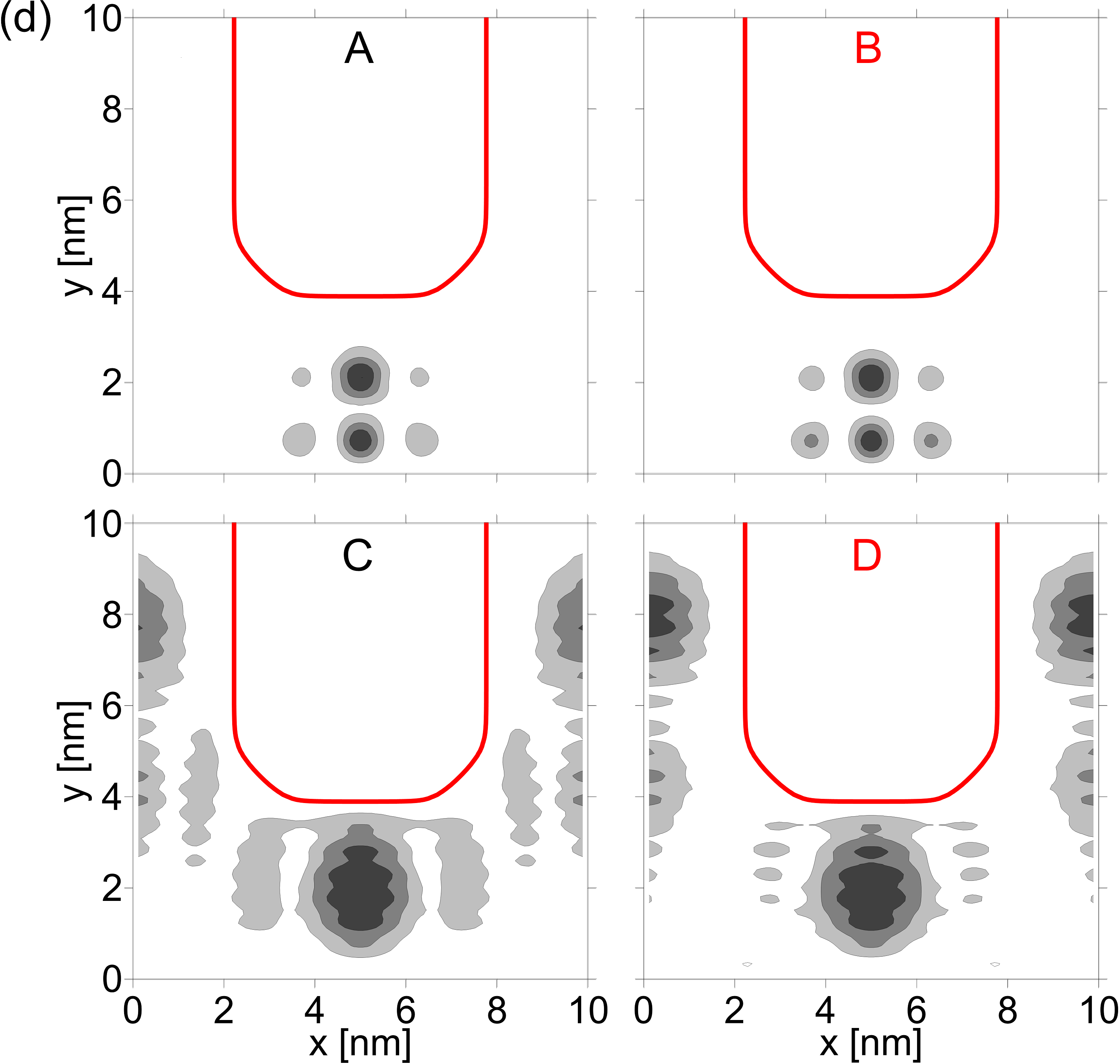}
     \end{minipage}
        \caption{Results for the lower $\sigma$ subband of the first pair - $\sigma_1$-lower, cf.~Fig.~\ref{fig:fitted_dispersions}. (a) Energy gap and average order parameter $\Delta$ as a function of the SC coupling parameter $J_0$. Solid lines correspond to the full calculation and dashed lines to the calculation where the mixing for miniband $4$ is turned off. Black lines are for the system with no phase difference and red lines for $d\phi=\pi/4$. (b) Energy dispersions for the case indicated as $A$ in (a). The insets show the normal state density at the marked $Q$-points. The energy gap (“gap”) and the energy shift for the miniband $4$ (“mb4”) are marked with vertical arrows. (c) As in (b), but for the case indicated as $B$ in (a). (d) The modulus of the order parameter $\Delta$ for the points indicated with the letters $A$-$D$, respectively, in (a). The thick red line shows the approximate effective ribbon/barrier limit - cf.~Fig.~\ref{fig:potential}.}
        \label{fig:s1_results}
\end{figure*}

In Fig.~\ref{fig:s1_results} we present results for the lowest-lying $\sigma$ subband (i.e.~the $\sigma_1$-lower subband - cf.~Fig.~\ref{fig:fitted_dispersions}). In this case, the Fermi level (zero energy in our energy scale) cuts across the miniband $2$ in the normal spectrum. The miniband $1$ lies in the Debye window $0 < E < E_C$. Minibands $3$ and $4$ lie in the negative counterpart of the window $-E_C < E < 0$, so their hole-like reflection will be present in the proper window itself. These minibands are presented in Fig.~\ref{fig:s1_results}(b), and they can in general contribute to the order parameter $\Delta$.

Two measures of superconductivity are shown in Fig.~\ref{fig:s1_results}(a) versus the $J_0$ coupling parameter: the energy gap (lower left scale) and the spatial average of the order parameter magnitude (upper right scale). Let us focus first on the solid lines, corresponding to the normal calculation. As expected, in the low $J_0$ limit, the gap is closed and there is no order parameter. If $J_0$ is increased, at some point the gap opens and simultaneously a non-zero order parameter is generated in the system, with a further increase in both parameters beyond the opening point. This switch from normal to SC state happens earlier in case of no phase difference, later for $d\phi=\pi/4$ and does not happen at all in the $J_0$ range considered in case of $d\phi=\pi/2$ [calculated, but omitted from Fig.~\ref{fig:s1_results}(a) for clarity]. This is in line with our previous results in Ref.~\cite{pasek2021band}, where the usual case was that the superconductivity was weakening as the phase difference was increased from zero to $\pi/2$. To study the SC state formation, we show the $Q$-dispersions for the two cases: in Fig.~\ref{fig:s1_results}(b) of the case marked as $A$ in (a), i.e.~$d\phi = 0$, $J_0 = 6$~meV, and in Fig.~\ref{fig:s1_results}(c) of the case marked as $B$ in (a), i.e.~$d\phi = \pi/4$, $J_0 = 7$~meV. For a detailed discussion, the reader can be referenced to the mentioned previous work, but there is one thing that strongly stands out. The high lying ($E{\sim}60$~meV) miniband $4$ is strongly shifted from the hole-like reflected normal state (marked mb4 in the figure and referenced this way in the following) to the corresponding upper-branch SC solution, as shown by the vertical black arrows.

Normally, the states far away from the Fermi level are only slightly shifted in energy by the SC coupling. To investigate further, we have redone the whole calculation, but with the mixing of mb4 manually turned off. This leads to the results presented in Fig.~\ref{fig:s1_results}(a) with the dashed lines. Already in the case of no phase difference, there is a significant quantitative impact of the mb4, both on the gap and on the average $\Delta$. However, the most interesting effect takes place for $d\phi=\pi/4$, as the $J_0$ value at which the gap is opened and $\Delta$ produced shifts from about $6.25$~meV to about $9.5$~meV. This marks a qualitative change in the state of matter introduced by the mb4 contribution, as the system would be in a normal state without it, while it is superconducting because of it.

This is an unusual situation, as the SC properties are usually thought to be decided by the characteristics of the normal system at, or close to, the Fermi level. We studied the mb4 to identify the cause of the phenomenon. The charge density of the normal state mb4 at $Q=0$ is shown in the upper inset of Fig.~\ref{fig:s1_results}(b), with the arrow and dot showing the corresponding state in the spectrum. One can see that the state is very strongly localised, lying wholly in the central constricted section of the primitive cell. On the other hand, the lower inset shows the density of the miniband $2$ at the point of cutting across the Fermi energy, which would normally affect the SC most strongly. The density of that state is distributed over the whole primitive cell, which is the typical scenario. This peculiar character of mb4 leads to an unusually strong contribution to the order parameter. The easiest to understand is the large shift in mb4 state itself. The $\kappa_{Q,n,Q,n,0}$ contact term is basically the integral over the cell of the $Q,n$ minibands’ density squared. As the integral of the density itself is normalised, the integral in question is the biggest when more density is squeezed into large local maxima.

Moreover, mb4 is nearly flat, which suggest that the wavefunction symmetry is not strongly affected by $Q$. This leads to relatively compatible symmetries of e.g.~$Q$ and $Q^\prime$ in the $\kappa_{Q,4,Q^\prime,4,0}$ term or $Q$ and $Q+dQ$ in the $\kappa_{Q,4,Q,4,dQ}$ term. The strong impact of the mb4 on the miniband $2$, leading to the enhancement and/or earlier opening of the gap, is more complicated because it also considers the relation between the symmetries of the two states in question. However, e.g.~in the particular case of the two densities shown in Fig.~\ref{fig:s1_results}(b), one can see that the symmetries are sufficiently fitted, with the state of miniband $2$ having relatively large density over the whole constricted part, where mb4 is localised. In contrast, the integral value would be minimal e.g.~if the miniband $2$ was localised mostly in the unconstricted part of the cell.

To sum up the above, two crucial factors were identified (flatness, strong localisation) that lead to the strong impact of mb4 on the gap and magnitude of the order parameter. The latter, however, does not say anything about the symmetry of $|\Delta(x,y)|$, which is shown in Fig.~\ref{fig:s1_results}(d) for the selected cases marked as $A$-$D$ in (a). One can easily see that the shapes of the order parameter in cases $A$ and $B$ are very similar to the shape of the mb4 in the upper inset of Fig.~\ref{fig:s1_results}(b). In contrast, the shapes of $|\Delta(x,y)|$ for $C$ and $D$ resemble much more the lower inset in Fig.~\ref{fig:s1_results}(b), which corresponds to the state of miniband $2$. We show here only $4$ cases, but the general shape of the order parameter changes very little in the SC range along each of the lines in Fig.~\ref{fig:s1_results}(a) as $J_0$ is changed, with only the overall magnitude being affected. In conclusion, another qualitative change connected to the mb4 is found, i.e.~the redefinition of the shape and spatial localisation of the order magnitude from the one corresponding to the miniband cutting across the Fermi level to one dominated by that high-lying miniband.

The phenomenon described above can be interpreted as an example of the Fano-Feshbach resonance \cite{salasnich2005condensate} coupling the minibands $2$ and $4$. In this kind of resonance, two processes interact, where one has a  discrete spectrum and the other - a continuous spectrum. In the context of multichannel superconductivity, the role of the latter was attributed to a deep-lying band,  and the role of the former to the band whose (locally flat) bottom is tuned to the Fermi level \cite{guidini2014band}. In our case, the miniband $2$ plays the role of the continuous element, and mb4 – which is nearly flat – has a quasi discrete spectrum. The difference here is that the chemical potential is not tuned with respect to mb4 and only the continuous element - miniband $2$ is cut across by Fermi level. On the contrary, mb4 is a high-lying energy band and so this coupling could be interpreted as a remote Fano resonance and the possibility of such remoteness is introduced by the characteristics of mb4 described in the previous paragraph.

\begin{figure}[htp]
         \includegraphics[width=0.496\textwidth]{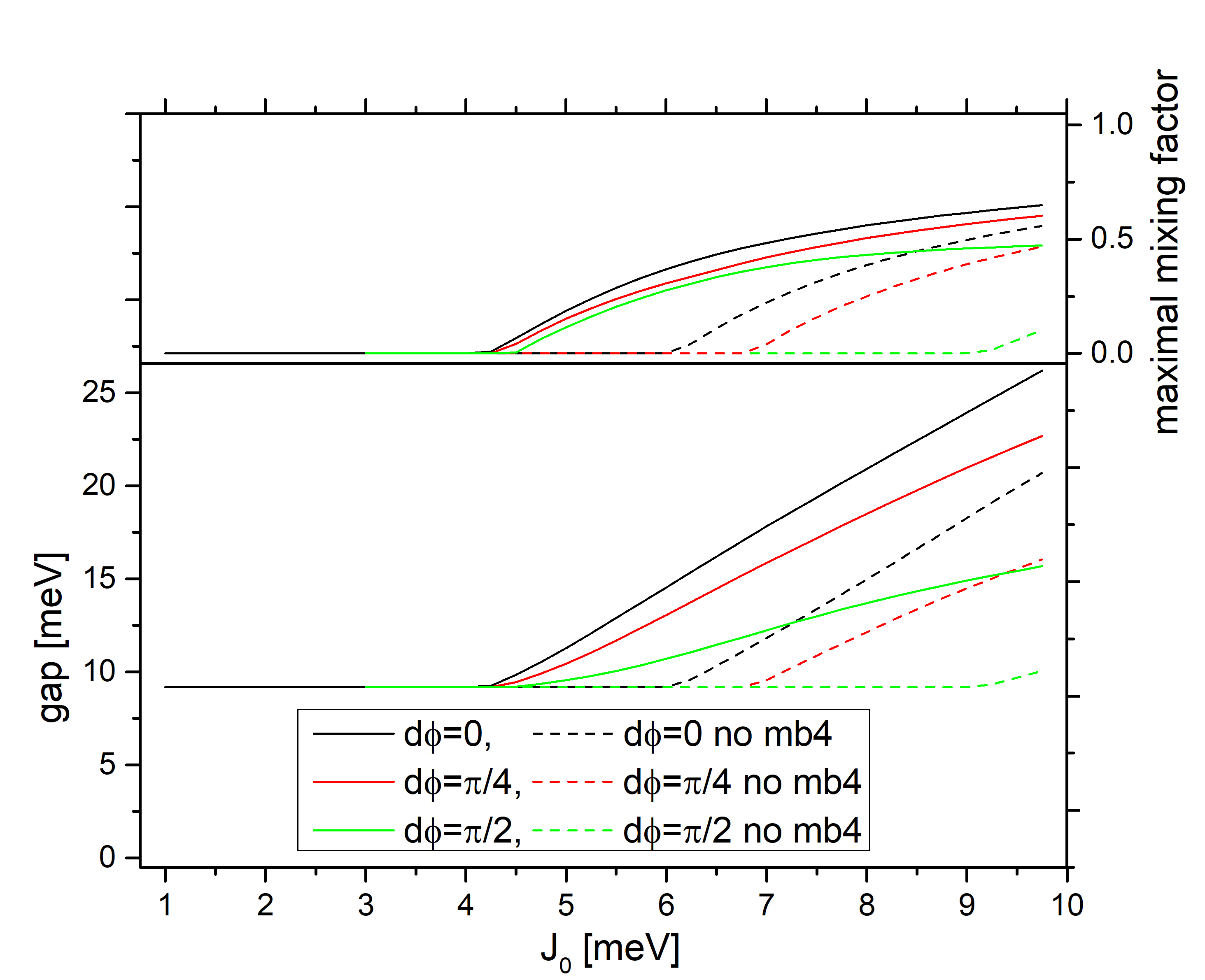}
        \caption{Results for $\sigma_1$-upper subband - cf.~Fig.~\ref{fig:fitted_dispersions}. Energy gap and maximal mixing factor as a function of the SC coupling parameter $J_0$. Black lines are for the system with no phase difference, red lines for $d\phi=\pi/4$ and green lines for $d\phi=\pi/2$.}
        \label{fig:s1b2_results}
\end{figure}

The energy gap and the maximal mixing factor for the $\sigma_1$-upper subband are shown in Fig.~\ref{fig:s1b2_results}. This is a similar version of Figs.~\ref{fig:s1_results}(a), with the exception that the maximal mixing parameter $\max_{Q,n} 1 - \left\lvert \left\lvert U_{Q,n} \right\rvert ^2- \left\lvert V_{Q,n} \right\rvert ^2 \right\rvert$ is used in place of the average order parameter, as the former is much faster to calculate and there is a strong relationship between the two. Namely, if one of them is zero, then the other is necessarily also zero. In the case of $\sigma_1$-upper subband, in the normal state, the Fermi level is not cut across by any miniband, it lies between them. This is a signature of an isolating (non-metallic) system and the gap, which is open for $J_0\rightarrow0$ is an insulating one. It stays the same for a range of small $J_0$ until finally begins to increase with $J_0$ and simultaneously non-zero mixing (and also non-zero $\Delta$) is generated in the system. Please note that the mixing increases slowly and even for quite large $J_0$ the mixing is not $\sim1$. This is because the gap has always two components here: the insulating as well as the SC-coupling one. Also here we can observe a significant impact of mb4 (compare the dashed and the solid lines). It leads to a much faster generation of SC and decreases the dependence on the phase difference. The latter effect is, again, the consequence of the flat mb4 being relatively insensitive of $Q$.

\begin{figure*}[htp]
     \centering
     \begin{minipage}[b]{0.496\textwidth}
         \centering
         \includegraphics[width=\textwidth]{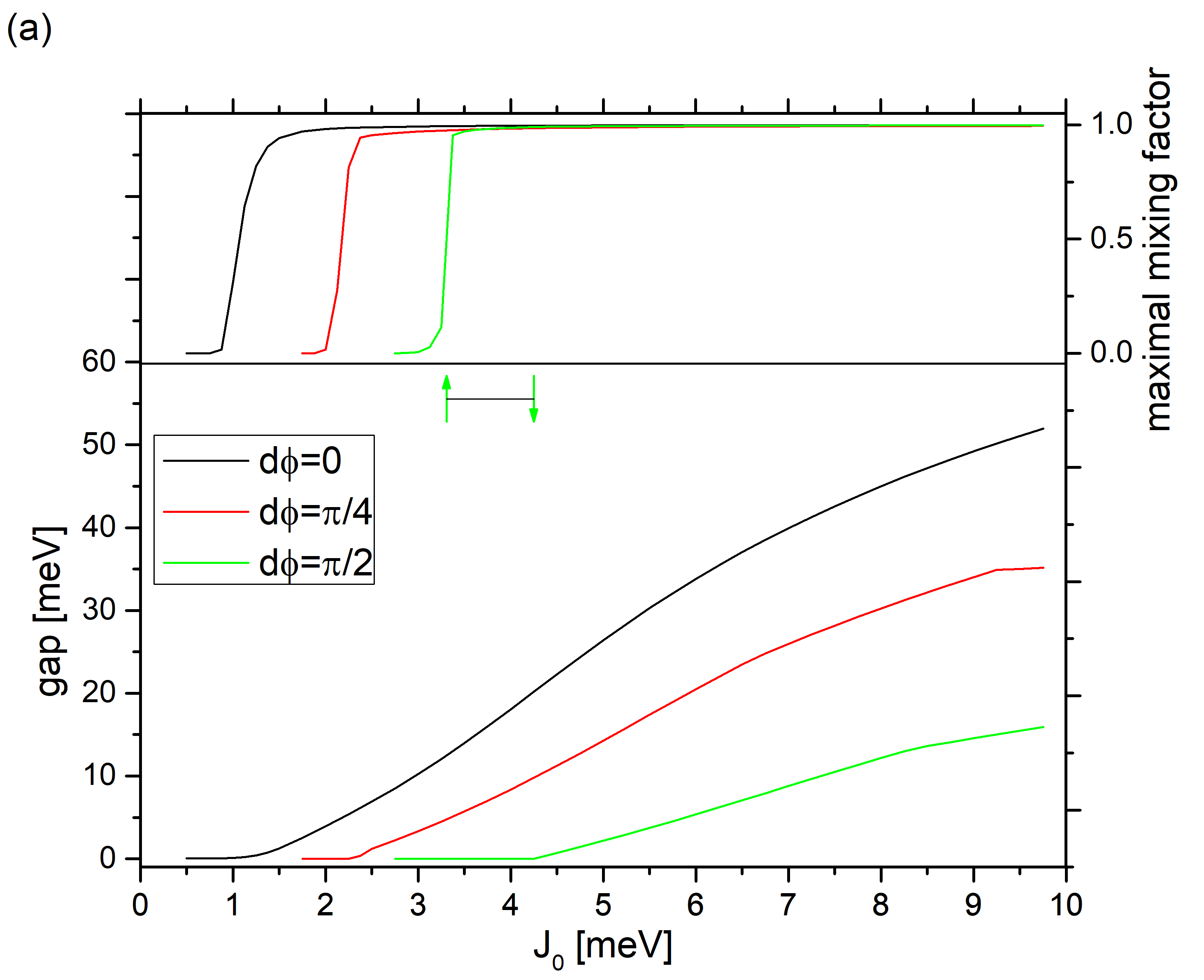}
     \end{minipage}
     \begin{minipage}[b]{0.496\textwidth}
         \centering
         \includegraphics[width=\textwidth]{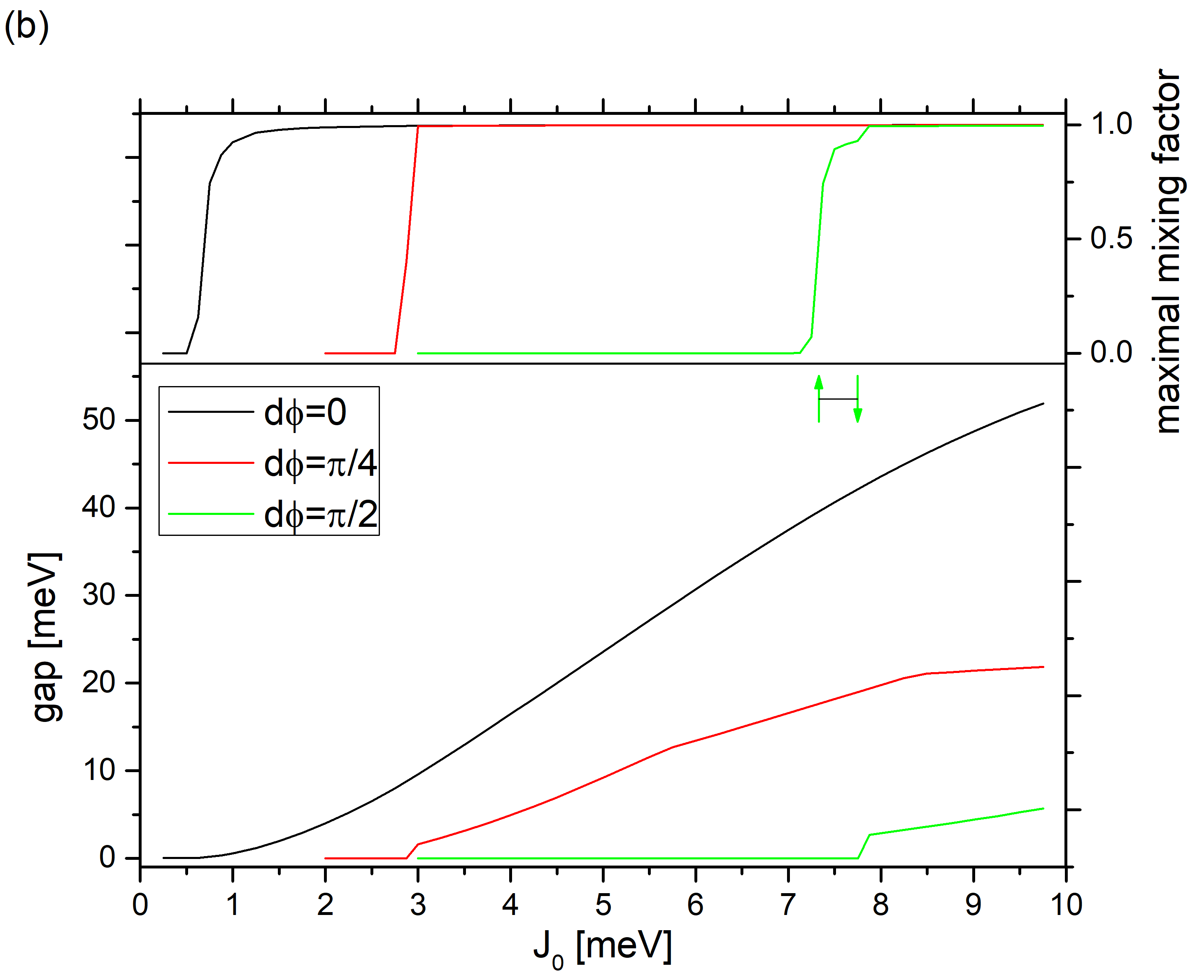}
     \end{minipage}
     \begin{minipage}[b]{0.496\textwidth}
         \centering
         \includegraphics[width=\textwidth]{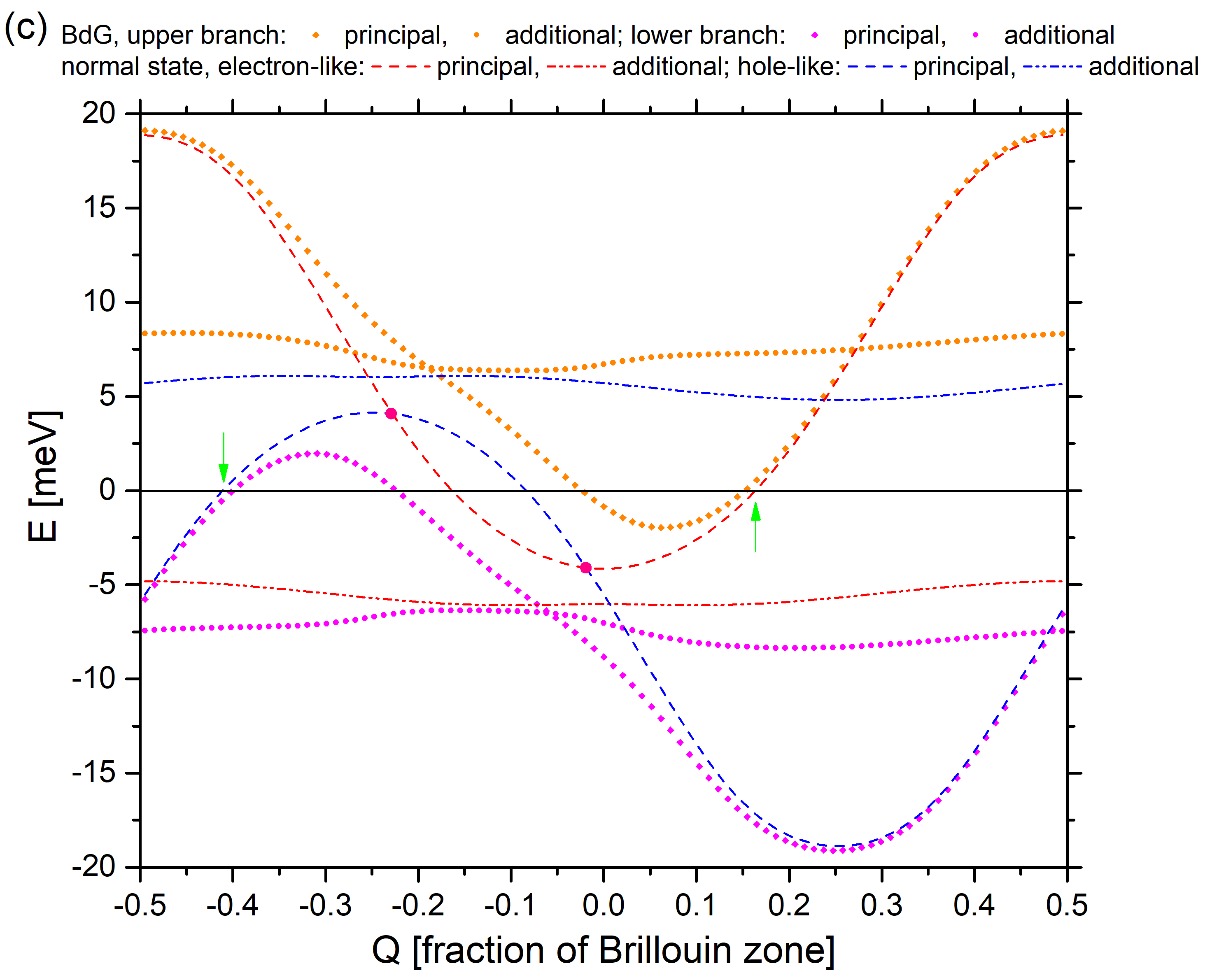}
     \end{minipage}
     \begin{minipage}[b]{0.496\textwidth}
         \centering
         \includegraphics[width=\textwidth]{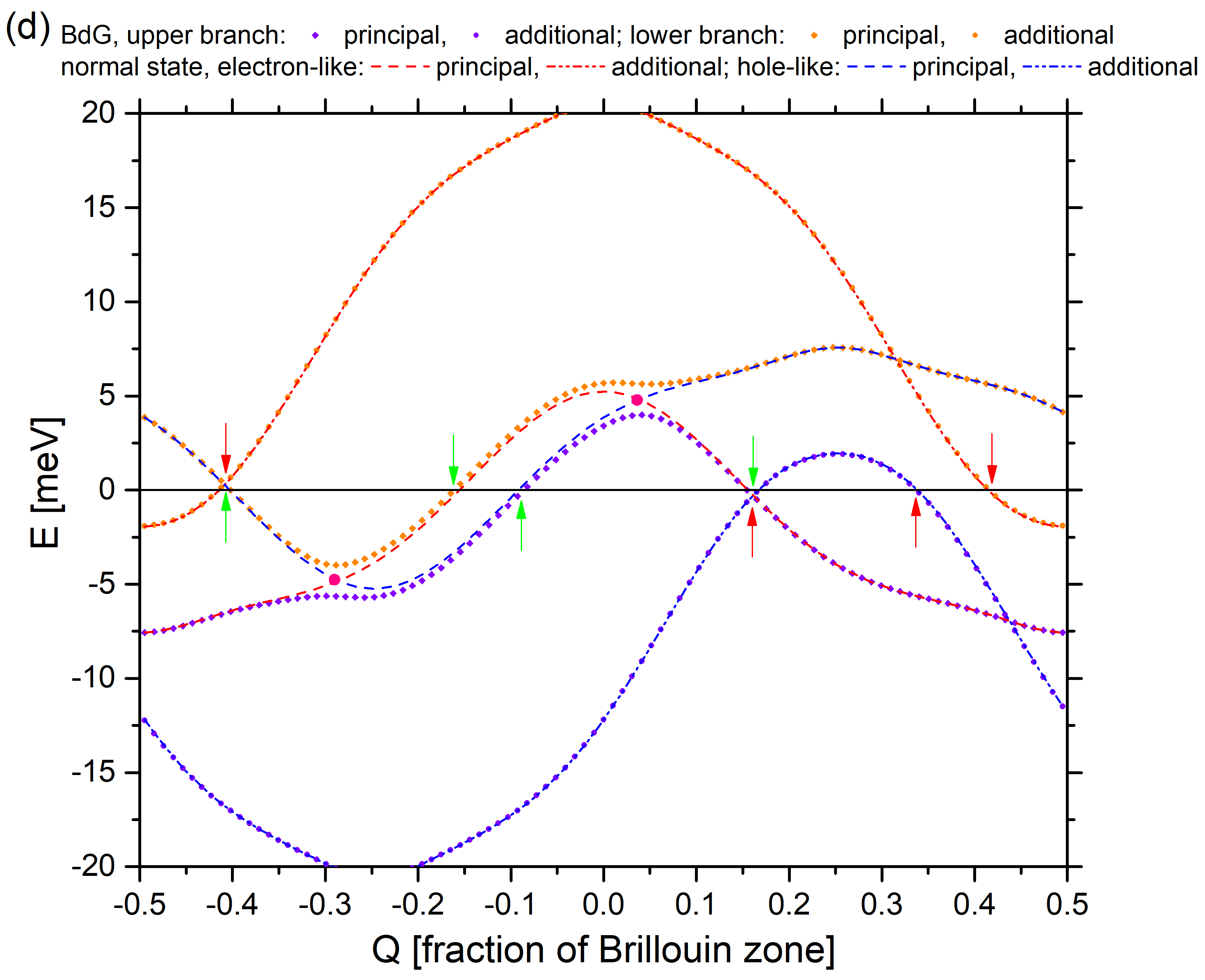}
     \end{minipage}
        \caption{Results for (a) $\sigma_2$-lower and (b) $\sigma_2$-upper subbands - cf.~Fig.~\ref{fig:fitted_dispersions}. Energy gap and maximal mixing factor as a function of the SC coupling parameter $J_0$. Black lines are for the system with no phase difference, red lines for $d\phi=\pi/4$, and green lines for $d\phi=\pi/2$. (c) $Q$-dispersion for the pseudogap case of $\sigma_2$-lower subband, for $d\phi=\pi/2$ and $J_0=4$~meV. (d) $Q$-dispersion for the pseudogap case of $\sigma_2$-upper subband, for $d\phi=\pi/2$ and $J_0=7.75$~meV.}
        \label{fig:s2_results}
\end{figure*}

\begin{figure*}[htp]
     \centering
     \begin{minipage}[b]{0.496\textwidth}
         \centering
         \includegraphics[width=\textwidth]{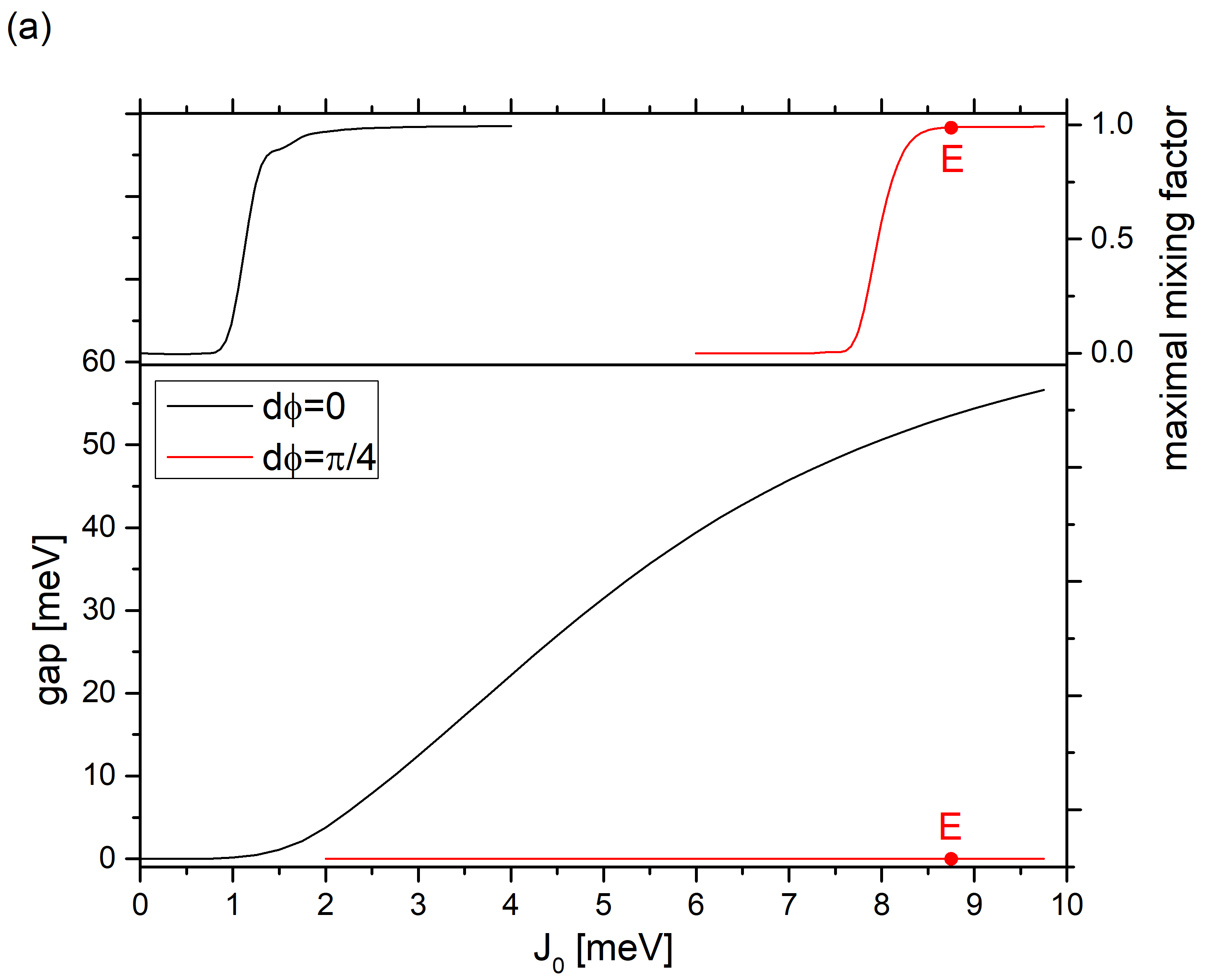}
     \end{minipage}
     \begin{minipage}[b]{0.496\textwidth}
         \centering
         \includegraphics[width=\textwidth]{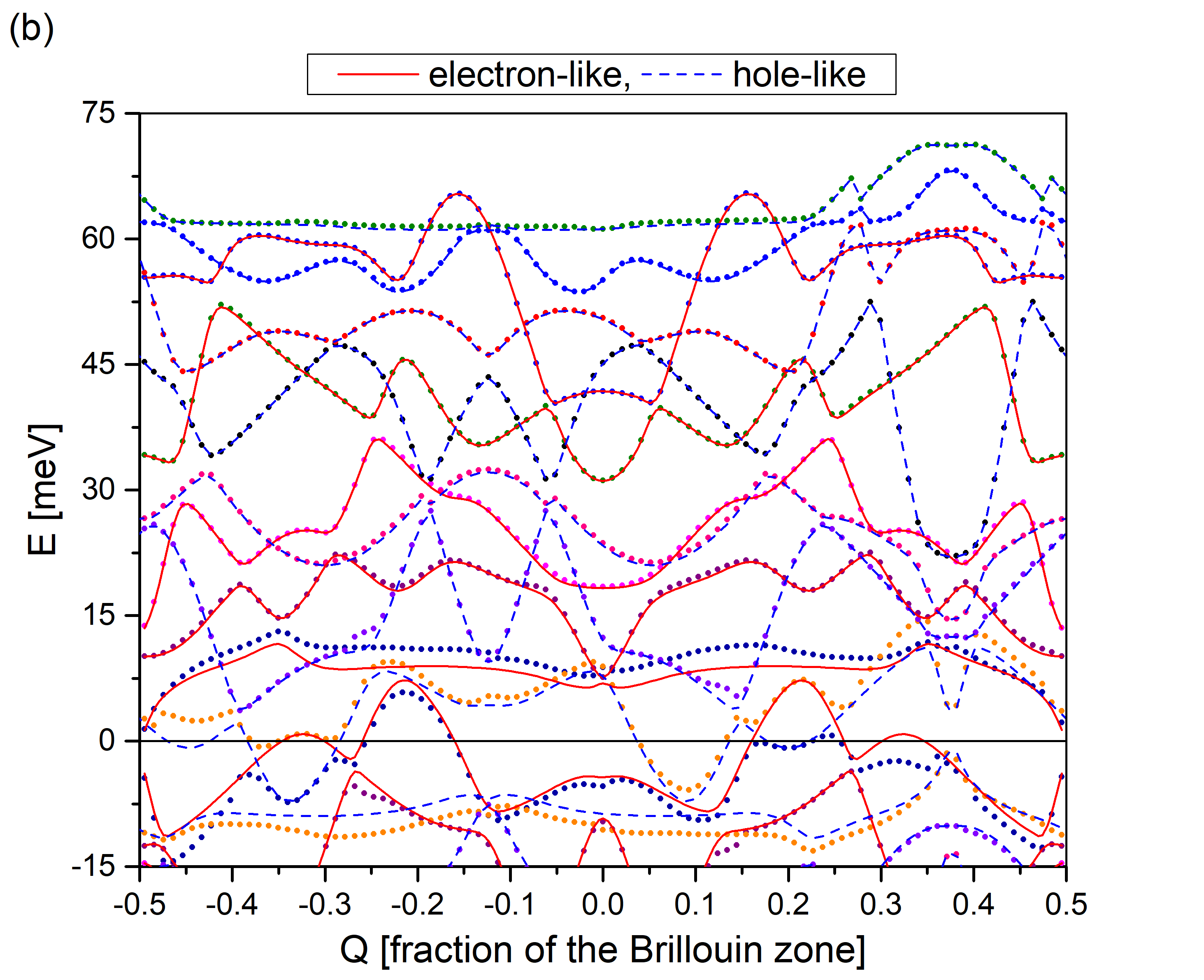}
     \end{minipage}
     \begin{minipage}[b]{0.496\textwidth}
         \centering
         \includegraphics[width=\textwidth]{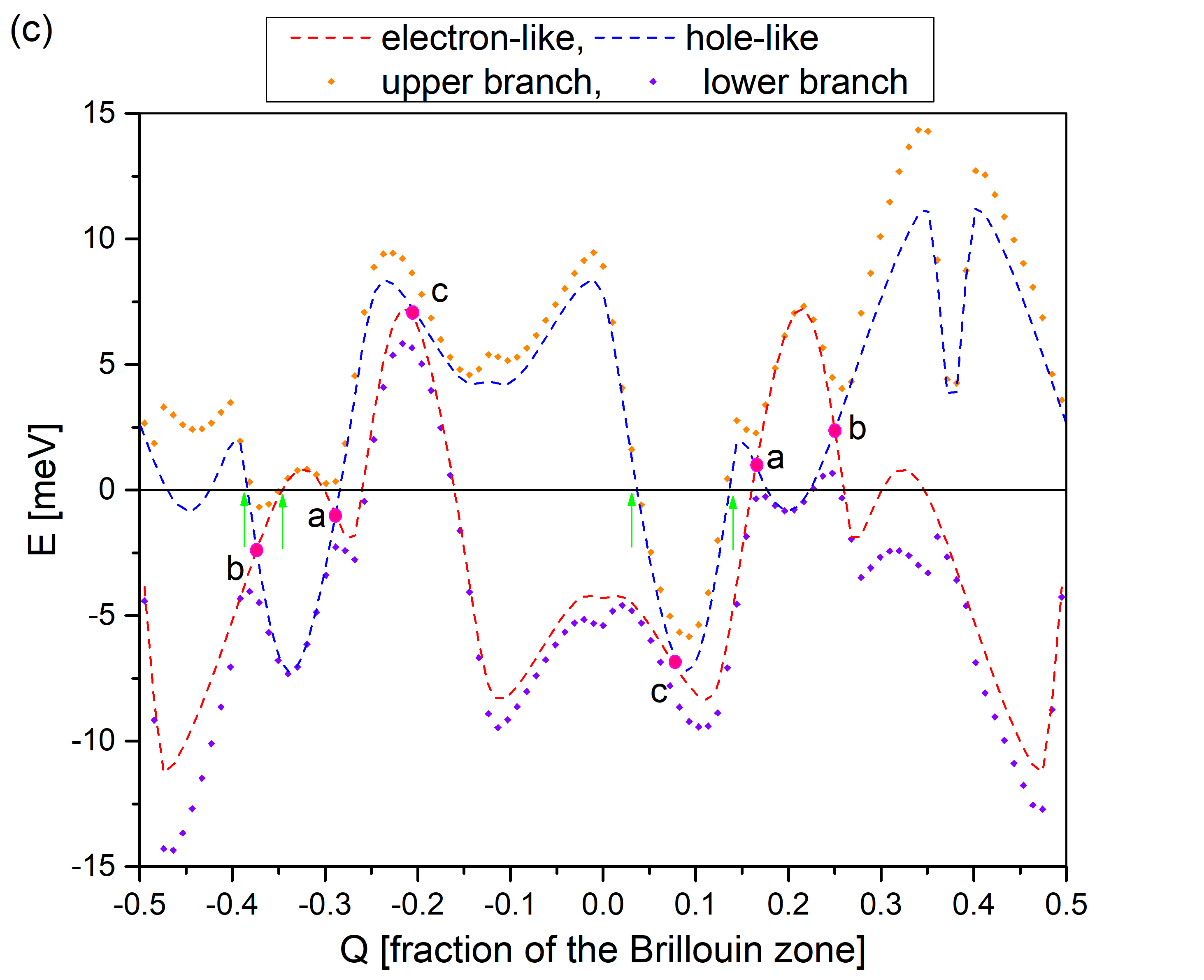}
     \end{minipage}
        \caption{(a-c) Results for the $\pi_1$ subband - cf.~Fig.~\ref{fig:fitted_dispersions}. (a) Energy gap and maximal mixing factor as a function of the SC coupling parameter $J_0$. Black lines are for the system with no phase difference and red lines for $d\phi=\pi/4$. (b) Single-particle energy dispersions for the case indicated as $E$ in (a). The Fermi level is set at $E=0$. (c) Details on the pseudogap formation in the $E$ case in (a). Only the miniband (red) cutting across the Fermi energy and its hole-like (blue) counterpart are shown, together with the upper and lower branches of the SC solution. The magenta dots show the crossing points between the electron and hole-like parts, around which the SC gaps/mixings are intensified. The green arrows show where the gap is closed by the upper branch.}
        \label{fig:p1_results}
\end{figure*}

The results for $\sigma_2$-lower and $\sigma_2$-upper subbands are shown in Figs.~\ref{fig:s2_results}(a) and (b), respectively. Here, the simple correlation between the gap and the mixing factor with $J_0$, observed in Figs.~\ref{fig:s1_results}(a) and \ref{fig:s1b2_results}, is no longer followed. One can see that in the case of $d\phi=\pi/2$ there is some small region between the point where the significant mixing is found (see the vertical pointing up arrows) and the one where the gap opens (see the vertical pointing down arrows). Within this region, in some parts of the Brillouin zone exists superconductivity (and open gap) whereas in other parts of the B.z. there is no SC coupling (and the gap is closed). These are the signatures of  state of matter called pseudogap.

In order to explain how it arises, it would be beneficial to imagine the simplest possible case first. Let us take into consideration only a single miniband cutting across the Fermi level, which is of simplest cosine-like $Q$-dispersion (with one maximum and one minimum in the first Brillouin zone), which we will call here the \textit{principal} miniband. When there is no phase difference in the system, the electron-like and hole-like parts cross\footnote{In this work the verb "to cross" is used to mark the electron- $\epsilon_{Q,n}$ and hole-like parts $-\epsilon_{Q+\dd{Q},n}$ crossing each other. "To cut across" is used to mark a miniband and $E_F$ crossing each other.} each other exactly at Fermi energy $(\epsilon_{Q,n}=0)$ at two opposite $Q$ values. It is well known\footnote{See e.g.~the detailed analysis of the usual dynamic in our previous work \cite{pasek2021band}.} that the crossing of the two dispersions is the place where the mixing (and hence the order parameter) is primarily generated. In this case, a slightest amount of nonzero $\Delta$ leads to energy shift which immediately opens the gap, as the energy shift is in fact equal to the gap. This is because \mbox{$E_{SC} = \pm \left ( \left \lvert \epsilon_{Q,n} \right\rvert + \left \lvert E_{\text{shift}} \right\rvert \right ) $}  and $\epsilon_{Q,n}=0$ at the point of crossing. However, in case of some non-zero $d\phi$, the crossing points move in energy to, respectively $\pm E_{\text{cross}}$, which means that $\epsilon_{Q,n}=0$ or $-\epsilon_{Q+\dd{Q},n}=0$ happen elsewhere in $Q$. For a relatively small $J_0$, when the mixing and the shift only starts to be generated, it exists only in the vicinity of the crossing points and is not strong enough to open the gap where the electron- or hole-part energy cuts across the Fermi level.

An in-depth curious reader may ask at this point how the existence of the small shift is even possible in the case that $E_{\text{shift}} < E_{\text{cross}}$, in which both the upper and the lower branches of the SC solution lie on the same (positive/negative) side of $E_F$. Does not the $U$-$V$ symmetry of the BdG solution lead to the cancelling out of the corresponding contributions? This was the origin of the usual $d\phi$-dependent SC collapse mechanism described in detail in our previous work \cite{pasek2021band}, where transition of one of the branches of the BdG solution to the other side of $E_F$ results in the total collapse of SC. Yes, that is generally the case, but with two caveats. Firstly, the contributions only cancel out exactly in the $T{\rightarrow}0$ limit [see the $1-2 F_D\left(E_{Q^\prime,\dd{Q},m},T\right)$ term in Eq.~\ref{eq:Delta_element}], while our simulation assumes a small but non-zero temperature. Secondly, there may be some \textit{additional} minibands relatively close to the Fermi level that may assist in the SC formation by not cutting across the $E_F$.

We can find an example illustrating our simple analysis in Fig.~\ref{fig:s2_results}(c), which presents the $Q$-dispersion for the pseudogap case of $\sigma_2$-lower subband, for $d\phi=\pi/2$ and $J_0=4$~meV. Here, the electron-like normal state of the \textit{principal} miniband cutting across the Fermi level is marked by red dashed line while the corresponding hole-like part is marked with blue dashed line. They have the general simple form of cosine functions. The resulting two crossing points between these lines are marked with magenta dots. The corresponding upper and lower BdG levels (the orange and violet diamond-shape points) are visibly shifted in the vicinity of the crossing points, marking the mixing and $\Delta$ generation, but not shifted enough for the upper and lower BdG solutions to lie on the opposite sides of $E_F$. In fact, the points where the electron- or hole-like parts cut across $E_F$ (green vertical arrows) have BdG and normal state energies nearly coinciding, generating only residual mixing. There is an \textit{additional} normal state miniband in the vicinity of $E_F$, which is shown as dashed-dotted lines: red for the electron- and blue for the hole-like parts. The corresponding BdG levels are marked by circle dots, and one can see that this \textit{additional} miniband is shifted (mixed) in most of the Brillouin zone, therefore it can act to assist in the $\Delta$ generation.

Another example of the pseudogap state is shown in Fig.~\ref{fig:s2_results}(d) for the case of $\sigma_2$-upper subband, for $d\phi=\pi/2$ and $J_0=7.75$~meV. This figure uses the same colour/symbol scheme as the previous one. Here, the \textit{principal} miniband has slightly more complicated shape, but still not very qualitatively far away from a cosine function, especially in respect to the minima/maxima points, resulting in two crossing points. The significant difference between this case and the (c) one is that the \textit{additional} minibands remain not shifted (unmixed) in the whole Brillouin zone. This means that it cannot aid in the $\Delta$ generation and that the effect is purely due to finite temperature.  Moreover, not only the \textit{principal} miniband cuts across the $E_F$ (see the green vertical arrows), as usual, but the \textit{additional} miniband does the same (see the red vertical arrows), meaning that in order for the gap to be open, a sufficient shift needs also to be generated by the \textit{principal} miniband in the \textit{additional} one.

Note that a small region of pseudogap of a similar kind of was found for $\sigma_1$ in Fig.~\ref{fig:s1_results}(a) for $d\phi = \pi/4$, for both mb4 included and excluded in the simulation.

Now let us focus on the results for the intra-intra model for the $\pi_1$ subband. A simple correlation between the open gap and nonzero mixing exists in the case of no phase difference [black lines in Fig.\ref{fig:p1_results}(a)]. However, the $d\phi=\pi/4$ case (red lines) is more interesting, as the gap is closed in the whole range of the $J_0$ parameter, but the mixing is present for $J_0 > 7.75$~meV. The $Q$-dispersion spectrum in the relevant point, marked $E$ in (a), is shown in Fig.~\ref{fig:p1_results}(b), where there is a plenitude of normal state minibands, of complicated shapes, overlapping energies and visibly interacting with one another. Consequently, there is also a multitude of BdG minibands, but one can see that in most cases the dots coincide with the corresponding lines – meaning no shift and no mixing is present for these states.

To have a clearer look at the situation, we present the subset of the dispersion of (b) in Fig.~\ref{fig:p1_results}(c). Here, only the electron- and hole-like parts of the miniband cut across by the Fermi level are shown, and their corresponding upper and lower branch BdG states.  The colour/symbol scheme is identical as in Fig.~\ref{fig:s2_results}(c,d), but only the points where the upper BdG branch cuts across $E_F$ are marked, for clarity.  In this case, the previously discussed pseudogap dynamic is still present, but there is also one peculiarity. Due to the more complicated shape of the normal state miniband, there are $3$ crossing point pairs, labelled as $a$, $b$ and $c$. The shift, mixing and $\Delta$ are mainly generated in each of these points. Each of the crossing pairs can have different behaviour and in this case near the crossing $a$ the upper and lower branches lie on opposite sides of $E_F$, which leads to a very robust mixing, while near the other two pairs, both branches lie on the same side of $E_F$, leading to only small temperature-dependent effect. Note that a similar case with multiple crossing points, which contribute very differently to the SC formation, was discussed for a pure $2$D model in NbSe$_2$ nanoribbon SL in our previous work \cite{pasek2021band}. The $\pi_1$ $d\phi=\pi/4$ case is the most significant example of pseudogap state that we have found in the results, which we link to this peculiarity.

\section{Discussion}

NbSe$_2$ nanosystem with a single constriction was investigated in \cite{flammia2018superconducting}. In that work, the longitudinal length of the nanoribbon was of a very different scale than the transverse width. This led to a Q1D normal state DOS being in fact composed of two DOS: the one corresponding to states localised in the nanoribbon and the ones localised in the constricted area (cf. Figs.~$2$ and $3$ in the mentioned work). In stark contrast, due to geometry of the NbSe$_2$ SL primitive cell in our previous work \cite{pasek2021band}, all normal state eigenstates were of essentially mixed $x$-$y$ and nanoribbon-constriction character. In the current study we found a remote resonance between miniband of a quasi-discrete spectrum and another with a continuous spectrum. The latter one is a state typically found in \cite{pasek2021band}. But the former one is fully localised in the constriction, similar to what was found in \cite{flammia2018superconducting}, in spite of the fact that here the shape of the primitive cell is similar to what it was in our previous work. The crucial difference here is the material (MgB$_2$ as opposed to NbSe$_2$) and the corresponding parameters of the model.

There is a large energy difference between the Fermi level and the subband energy at the $\Gamma$ point, even for the closest-lying $\sigma_1$ ($267.0$ meV cf. Table \ref{tab:eff_masses}) which means that the minibands close to the Fermi level are highly excited ones, counting from the vertex of the dispersion parabola. This may manifest as a localisation of the state of simple symmetry in a much smaller area of the constricted part [cf. the top inset -- for mb4 -- in Fig.~\ref{fig:s1_results}(b)]. For comparison, in \cite{pasek2021band}, the states around the Fermi level were the several first ones, counting from the vertex of the parabola i.e. the low excited ones. As a consequence, there was no possibility of encountering the strongly localised states in the Debye window and hence no possibility of the resonance to happen.

\section{Further developments of the model}

Because of the enormous computational complexity, we started to build our model ground up. This work shows the results for the intra-subband intra-miniband version of the Anderson model.

The first inter-\textit{subband} extension should include two sigma subband pairs $(\sigma_{1l} , \sigma_{1u} , \sigma_{2l} , \sigma_{2u})$ and one $\pi$ subband $(\pi_1)$. Please note that this is the minimal model that allows to study the inter-subband coupling between the $\sigma$ \textit{l}-\textit{u} subbands from the same pair, the \textit{l}-\textit{l} and \textit{u}-\textit{u} mixing between the pairs, the \textit{l}-\textit{u} \textit{1}-\textit{2} mixing, and the $\sigma$-$\pi$ couplings. Each of these can introduce new dynamics to the SC formation.

The multitude of minibands and their energy overlap in Fig.~\ref{fig:p1_results}(b) is a strong argument for expanding the model to the inter-\textit{miniband} case, as a corresponding multitude of couplings is omitted. On the other hand, it was shown in \cite{salasnich2019screening} that the inter-subband coupling can minimise the pseudogap regime of parameters, which was originally caused by quantum fluctuations. In our work, the pseudogap arises as a consequence of the non-zero phase difference, but we still anticipate the mitigating effect of the inter-subband coupling.

In concussion, we see great utility and significance in expanding the Anderson model beyond its usual limitations, which would be the first goal of our project beyond the initial results we already have obtained.

Please note that our fit approach allows in general to use non-parabolic dispersions fitted to the data of Fig.~5 of \cite{bekaert2017free}, as long as the dispersion function is even in $k_{x(y)}$. This is because the kinetic energy operator can be expanded as the series in $k_{x(y)}^2$, and the mesh functions are the eigenfunctions for each of the terms of that series. Recognising the fact that more realistic fits are possible this way, we decided to use the simplest parabolic approach in this work in order to exclude additional effects coming from that and focus on the ones coming from the coupling of the subbands.

\section{Conclusions}
In this work, a parabolic effective mass model was developed for few-monolayer MgB$_2$ which allows to simulate confined superconductivity in Q1D structures. An example of application of the model was given with a simulation of a constricted SL in the scope of minimal intra-subband intra-miniband Anderson approximation. A remote Fano-Feshbach resonance was identified, which arises due to interplay between the constriction geometry and the large-$k$ components of the eigenstates close to the Fermi level. Also, it was shown how the system may exists in a pseudogap phase as a consequence of the non-zero phase difference imposition.

\section{Acknowledgements}
This work was supported by the U.S.~Office of Naval Research [award number N62909-19-1-2130]. We acknowledge computational support by LaSCADo [FAPESP – grant number 2010/50646-6] as well institutional support by Faculdade de Tecnologia - Unicamp, Brazil.

\bibliographystyle{model1-num-names}

\bibliography{bibliography.bib}

\end{document}